\documentclass[aps,pre,floatfix,nofootinbib,showpacs,twocolumn]{revtex4}
\usepackage{graphicx}
\usepackage{amsmath}
\usepackage{amssymb}
\begin{document}
\date{\today}
\author{Ramandeep S. Johal}
\affiliation{Department of Physics, Indian Institute of Science Education and Research Mohali,\\
Transit Campus: MGSIPA Complex, Sector 26, Chandigarh 160019, India}
\draft
\title{Quantum heat engines and nonequilibrium temperature}

\begin{abstract}
A pair of two-level systems initially prepared in different thermal states and coupled to
an external reversible work source, do not in general reach a common temperature at the end of a unitary work
extraction process. We define an effective temperature for the final nonequilibrium 
but passive state of the bipartite quantum system and analyse its properties.
\end{abstract}
\pacs{ 05.30.-d, 05.70.Ln, 05.70-a}
\maketitle
%
\section{Introduction}

Consider the thermodynamic problem of work extraction \cite{Callen, Landau}
from two systems at different temperatures $T_1$ and $T_2$
(let $T_1 >  T_2$) by coupling them with a reversible work source. It is assumed
that internal energy of each system is $U_i = C_i T_i$, where 
$C_i$ is independent of temperature. The process of work extraction stops when
 the two systems reach a common final temperature $T_f$.
Work performed is given by the difference of initial and final
energies:
\begin{equation}
W_0 = C_1 T_1 + C_2 T_2 - (C_1 + C_2)T_f.
\label{wc}
\end{equation}
Now to extract maximal work, the process is assumed to be thermally isolated
in which thermodynamic entropy of the total bipartite system is preserved. 
 This criterion yields the value of the final temperature
as $T_f =(T_1)^{\xi/(1 + \xi)} (T_2)^{1/(1 + \xi)}$, where  $\xi = C_1/C_2$..

One can discuss a cyclic process which proceeds
in the following two steps: i) the two systems prepared as above and
coupled to a reversible work source, are used to extract an amount of work given in Eq. (\ref{wc})
whence the systems reach a common final temperature; ii) the systems are
then brought back to their initial states by separating them from the 
work source and making contact with thermal baths at $T_1$ and $T_2$
respectively. In the second step, the system 1 absorbs heat from the hotter
bath and system 2 rejects some heat to the cold bath. The efficiency of this cyclic process is 
\begin{equation}
 \eta(\xi,\theta) = 1 + \frac{1}{\xi} \frac{\theta - \theta^{1/(1+\xi)}}{1-\theta^{1/(1+\xi)}},
\label{effc}
\end{equation}
where $\theta = T_2 /T_1$.
This system behaves very similar to a cycle discussed by Leff \cite{Leff}
which is made up of a sequence of infinitesimal Carnot cycles
and where both the heat source and the sink have a finite heat capacity.

The problem of work extraction  has also been addressed from a
quantum mechanical point of view \cite{Hastapoulus76, Scully2001, AA2004}. 
 Although the possibility of a quantum heat engine
and validity of thermodynamic bounds has been
recognised since 1950s \cite{Scovil}, the recent developments in nanotechnology and quantum
information processing have contributed to enhanced interest
in quantum thermodynamic machines \cite{Alicki2004, Mahler2004}.
Alongside, such models provide insight into fundamental questions
about thermodynamics such as Maxwell's demon and universality of the second law \cite{Lloyd1997, Vedral2008, Ueda2009}.
Many models employ few-level quantum systems as the working medium, 
such as quantum harmonic
oscillators, spin-systems, particle-in-box and so on 
\cite{Geva1992, Opatrny2002, Hua2002, Kieu2004, Quan2007}.
Usually the cycle is a quantum generalization of the well-known classical
Carnot, Otto, Brayton heat engines which follow four-step cycles.
In another class of models, instead of the two classical or macroscopic systems 
as discussed in the preceding paragraph, 
one can form a two-step engine using two quantum systems \cite{AA2004}.
Recently, such a quantum heat engine employing two two-level systems (TLS)
was discussed and implications of the optimization of work on the structure
of the engine were highlighted \cite{AJM2008}. 

However, quantum engines being small systems, the validity of thermodynamic
behaviour is not guaranteed. For instance, after work extraction in the latter
class of models, the two systems may not reach  mutual equilibrium. 
In this paper, we further discuss the two-step model for work extraction
using two TLS, focusing on the final passive state (which is in general
a nonequilibrium state) from a thermodynamic perspective.
We define an effective temperature for this state and analyse its properties.
The paper is organised as follows.
In section II, we introduce the model of quantum heat engine. In section IIA, 
the temperatures of subsystems are evaluated; the validity of thermodynamic definitions
is enforced by deriving the specific heats of subsystems in section IIB.
Section III proposes a definition for effective temperature of the composite
system, which is calculated explicitly in different regimes of parameter values.
We also compare some of the other definitions in literature for effective
temperature of nonequilibrium systems, in section IV.
Concluding ideas are given in section V.
\section{Quantum model for work extraction}
Consider two TLS labeled $R$ and $S$ with hamiltonians $H_R$ and $H_S$,
 prepared in thermal states  $\rho_R$ and $\rho_S $ corresponding
to temperatures $T_1$ and $T_2$. 
The hamiltonian of the total system is $H=H_R \otimes I + I\otimes H_S$.
The initial state of the composite system is $\rho_{\rm in} = \rho_R \otimes \rho_S$.
The eigenvalues of $H$ are $\{0, a_2, a_1, a_1+a_2 \}$  given that energy eigenvalues of
$H_R$ and $H_S$ are $(0,a_1)$ and $(0,a_2)$, respectively. The eigenvalues
of the initial density matrix are $\{r_1s_1, r_1s_2, r_2s_1, r_2 s_2 \}$.
Here the probability to find each system in its
excited state is
\begin{equation}
r_2 = \frac{1}{1+e^{a_1/T_1}}, \qquad s_2 = \frac{1}{1+e^{a_2/T_2}}, 
\label{}
\end{equation}
with ground state probabilities being $r_1 = (1-r_2)$ and $s_1 = (1-s_2)$.
We set Boltzmann constant $k_{\rm B} = 1$.

The initial mean energy of the composite system is $U = a_1 r_2 + a_2 s_2$.
Let us for concreteness choose, $a_1 > a_2$. 
Within the approach based on quantum thermodynamics, the process of work extraction
is a unitary process
which preserves not only the magnitude of the entropy but also all  
eigenvalues of the density matrix describing the state of the system.
It has been shown in earlier
works \cite{Hastapoulus76, AA2004, AJM2008} that under such a process, the state which
corresponds to a \textit{minimum} value of the final energy is $\rho_f = \rho_S \otimes \rho_R $,
with eigenvalues $\{r_1s_1, r_2s_1, r_1s_2, r_2 s_2 \}$.
Effectively, it means that the two systems exchange or swap their initial probability distributions
in the final state.
In other words, work performed is maximum if $U^{'} = a_1 s_2 + a_2 r_2$
and is given by 
\begin{equation}
{\cal W}(a_1,a_2) = U^{'} - U = (a_1-a_2)(s_2-r_2). 
\label{work}
\end{equation}
Net work is extracted if ${\cal W}<0$ which 
requires the following condition:
\begin{equation}
s_2 < r_2  \implies \frac{T_1}{T_2} > \frac{a_1}{a_2}.
\label{negativework}
\end{equation}
The efficiency of this engine is $\eta = 1-\frac{a_2}{a_1}$, which is independent
of temperature and its upper bound is Carnot value.
\subsection{Temperatures of subsystems after work extraction}
Now we study temperatures in the final state. 
After work, the mean energy of
subsystem 1 is $U^{'}_{1} = a_1 s_2$,
and $U^{'}_{2} = a_2 r_2$.
Let us consider two such set-ups specified by the pair of energy parameters 
$(a_1, a_2)$ and $(a_1 + d a_1, a_2 + d a_2)$. Comparing the
final states after work extraction, the change in energy of subsystem 1 
is 
\begin{equation}
dU^{'}_{1}=  s_2  da_1  + a_1 \frac{d s_2}{d a_2} da_2.
\label{dbu}
\end{equation}
We follow the standard interpretation of work as the change in mean energy due to shift in energy levels,
at constant probabilities \cite{Reif, Alicki1979, Kieu2004}. Similarly, heat is defined to  be the change in
mean energy when the energy levels stay fixed, but probability of occupation
changes. Thus the heat contribution for system 1 is given by
\begin{equation}
d Q^{'}_{1} = a_1  \frac{d s_2}{d a_2}  da_2.
\label{heat1}
\end{equation}
Similarly for subsystem 2, we have
\begin{equation}
d  Q^{'}_{2}= a_2 \frac{d r_2}{d a_1}  da_1.
\label{heat2}
\end{equation}
Let us now study entropy of each subsystem. In the initial state,
the entropy of subsystems are given by ${S}_1 = -(r_1\ln r_1 + r_2\ln r_2)$
and ${S}_2 = -(s_1\ln s_1 + s_2\ln s_2)$ respectively.
After work,  due to exchange of probabilities between the subsystems, we have
$S^{'}_{1} = S_2, S^{'}_{2} = S_1$.
Thus for subsystem, say 1,  
the change in entropy of the final state under a variation of the parameter $a_2$ is
\begin{eqnarray}
d S^{'}_{1} & = & dS_2  \\
& = & \frac{a_2}{T_2} \frac{d s_2}{d a_2}  da_2.
\end{eqnarray}
Now we evaluate the final temperature of system 1 as
\begin{equation} 
{T}_{1}^{'} \equiv \frac{d Q^{'}_{1}} {d S^{'}_{1}}=T_2 \frac{a_1}{a_2}.
\label{tp1}
\end{equation}
Similarly, we obtain for system 2 
\begin{equation}
T_{2}^{'} = T_1\frac{a_2}{a_1}.
\end{equation}
These values of temperatures are precisely which may be
obtained directly from the final probability distributions of the TLS,
because a TLS can always be assigned an effective temperature.

Using Eq. (\ref{negativework}), it can be seen that after work extraction, 
the hotter subsystem 1 cools
down (${T}_{1}^{'} < T_1$), where as the relatively cold subsystem 2 now has a 
higher temperature (${T}_{2}^{'} > T_2$).
Note that the sign of difference $(T_{1}^{'} - T_{2}^{'})$ is not determined;
it is possible to have $(T_{1}^{'} < T_{2}^{'})$. But this does not violate
the second law, because the condition (\ref{negativework})
also ensures that energy flows from the hot to the cold system. Thus the change
in energy of system 1, $\Delta U_1 = a_1(s_2 -r_2) <0 $ and the
corresponding change in system 2 is $\Delta U_2 = a_1(r_2 -s_2) >0 $.
%
\subsection{Heat capacity of subsystems}
The canonical heat capacity of subsystem 1 in the final state
is related to the fluctuations of energy in a well-known way \cite{Landau}.
However, heat capacity may also be evaluated as follows.
Consider the final temperature as function of $a_1$ and $a_2$ (Eq. (\ref{tp1})).
Then a change in temperature resulting from a variation in 
these parameters is
\begin{equation}
d T^{'}_{1} =\frac{T_2}{a_2} d a_1   -\frac{ a_1 T_2 }{(a_2)^2} d a_2.
\end{equation}
Then keeping $a_1$ fixed (which is equivalent to
keeping volume of subsystem 1 fixed, because change in $a_1$ for subsystem 1
in the final state is interpreted as work, see Eq. (\ref{dbu})), the heat capacity
(at constant volume) in the final state of system 1 is
\begin{eqnarray}
C_{1}^{'} & = & \left(\frac{\partial U^{'}_{1}}{\partial T^{'}_{1}} \right)_{a_1} \\
          & = &  C_2, 
\end{eqnarray}
where we have used the following identity 
 \begin{equation}
 -\frac{d s_2 }{d a_2} = \frac{C_2 T_2}{(a_2)^2}.
\label{ds2a2}
 \end{equation}
Here $C_2$ is the canonical heat capacity of the subsystem $2$ in its initial state at
temperature $T_2$. Similarly, we get the result $C_{2}^{'} = C_1$.
Thus upon swap-transformation, the specific heats of the two
subsystems also get exchanged.

To recapitulate, the standard thermodynamic process in which two 
macroscopic bodies at different temperatures are coupled to a work source,
the final temperatures of the two
bodies are said to become equal.  In the quantum
framework, the subsystems in general do not reach mutual thermal equilibrium. 
In the next section, we ask: can
the whole bipartite system be characterised by a global effective temperature
in the final state, even though it is a nonequilibrium state with
subsystems at different temperatures  ?
\section{ 'Temperature' for the bipartite system}
For subsystem $i$, we observed in the previous section that temperature
can be defined thermodynamically. In this section, we extend the
thermodynamic definition to the nonequilibrium final state 
of the composite system.

For convenience, we define ${a_2}/{a_1} = \nu$.
So the final temperatures are rewritten as:
$T_{1}^{'} = T_2/\nu$ and  $T_{2}^{'} = T_1 \nu$.
Thus for given reservoir temperatures ($T_1, T_2$),
the final temperatures of subsystems is determined by
a single parameter $\nu$, which is also related to
the efficiency of the engine $\nu = 1 - \eta$. Consider different final and 
initial states which are characterized by the same
parameter $\nu$, but which may yield different amounts
of work.

At a given value of $\nu$, the changes in $a_1$ and $a_2$ 
are related as 
\begin{equation}
da_2 = \nu da_1. 
\label{da2da1}
\end{equation} 
Thus the heat exchanged by system 1 in such a process
can be rewritten  from Eq. (\ref{heat1}) as
\begin{eqnarray}
d Q^{'}_{1} &=& a_1  \frac{d s_2}{d a_2} \nu  da_1  \\
           &=& a_2 \frac{d s_2}{d a_2}  da_1.
\label{}
\end{eqnarray}
Also, Eq. (\ref{heat2}) yields $d Q^{'}_{2}$ 
Then the total heat exhanged by bipartite system is 
$ d {Q^{'}} = d Q^{'}_{1} + d Q^{'}_{2}$.

Similarly, the von-Neumann entropy of the bipartite system is the sum
of subsystem entropies, $S^{'} ={S}^{'}_{1} + {S}^{'}_{2} =  S_1 + S_2$, 
and the variation in total entropy is 
\begin{equation}
d S^{'}  = \left[ \left( \frac{d {S}_1}
{d a_1} \right)  + 
\nu  \left( \frac{d{S}_2}
{d a_2} \right)\right] d a_1, 
\end{equation}
at a given $\nu$, using Eq. (\ref{da2da1}).
Then we define the effective temperature as the ratio of heat variation to the 
entropy variation, $T \equiv  \left(\frac{d {Q^{'}}}{d {S^{'}}}\right)$,
yielding
\begin{equation}
T = \frac{a_2\left(\frac{d s_2}{d a_2} + \frac{d r_2}{d a_1}  \right)}
 {\left( \frac{(a_2)^2}{a_1 T_2} \frac{d s_2}{d a_2} 
+ \frac{a_1}{T_1} \frac{d r_2}{d a_1} \right)}.
\label{tet}
\end{equation}
Using Eq. (\ref{ds2a2}) and $-\frac{d r_2 }{d a_1} = \frac{C_1 T_1}{(a_1)^2}$, we finally get
 \begin{equation}
  T = \frac{C_2 {T}_{1}^{'} + C_1 {T}_{2}^{'} }{(C_1 + C_2)}.
\label{tfeta} 
\end{equation}
The above formula is the main result of the present paper.
It resembles the thermodynamic expression if the two systems at
temperatures ${T}_{1}^{'}$ and ${T}_{2}^{'}$ with constant heat 
capacities $C_2$ and $C_1$ respectively, come to a common
temperature $T$, without doing any work, (see Eq. (\ref{wc})).
%
Now we evaluate the effective temperature for the 
case of two TLS, and discuss its features.
The canonical heat capacity of a TLS is given by the well known expression
\begin{equation}
C_i = \left(\frac{a_i}{T_i} \right)^2 \frac{\exp[a_i/T_i]}{(1+\exp[a_i/T_i])^2}.
\end{equation}
We first discuss the limit when $a_i/T_i=x  \ll 1$. Then $C_i(x) \sim x^2/4$
and the ratio $C_1/C_2  \equiv \xi \to  (\theta/\nu)^2$. Thus the efficiency
is given by $ \eta = 1-\frac{\theta}{\sqrt{\xi}}$. The effective
temperature in this regime is 
\begin{equation}
\frac{T}{T_1} = \frac{\sqrt{\xi}}{(1+\xi)} (1+ \theta).
\label{tmin}
\end{equation}
At Carnot limit, $\xi \to 1$ and $T/T_1 = (1+\theta)/2$.
However, note that this formula 
holds in general also, because Carnot limit implies $a_1/T_1 \to a_2/T_2$ and so
$C_1/C_2  \to 1$. Here the extracted work is vanishingly small and the final
temperature is expected to be $(T_1 + T_2)/2$.

At the other extreme, for $\xi \to \theta^2$, we have  
$T/T_1 = \frac{(\theta + \theta^2)}{(1 + \theta^2)}$.

In other words, when $a_2 \to a_1$,  ${\cal W} \to 0$ (see Eq. (\ref{work})).
Then ${{T}_1}^{\prime} = T_2$
and ${{T}_2}^{\prime} = T_1$ in this limit. In this case, the effective temperature is
simplified to
\begin{equation}
 \frac{T}{T_1} = \frac{(\theta + \xi)}{(1+\xi)}.
\label{tc1c2}
\end{equation}
Finally, we make the following observations:

i) it is interesting to note that the effective temperature is a weighted average 
of the subsystem temperatures after work extraction.

ii) the overall temperature
is the same as the subsystem temperature when the latter are
also equal to each other. This corresponds to $\nu = \sqrt{\theta}$,
which implies the well-known Curzon-Ahlborn efficiency \cite{Curzon1975}.

iii) At the global maximum of work, the conditions $\frac{\partial {\cal W}}{\partial a_1} = 0$
and  $\frac{\partial {\cal W}}{\partial a_2} = 0$ determine optimal values $a_{1}^{*}$ and $a_{2}^{*}$
as well as the condition
\begin{equation}
\frac{d r_2 }{d a_1}= \frac{d s_2}{da_2},
\label{extreme}
\end{equation}
holds. So we have $\nu^* =  \sqrt{\theta / \xi^*}$, where now  $\xi^*$ is determined from
using optimal values $a_{1}^{*}$ and $a_{2}^{*}$ . Then it follows from Eq. (\ref{tet}) that
\begin{equation}
{T}^* =  \frac{2{{T}_1}^{'} {{T}_2}^{'}} {{{T}_1}^{'}+{{T}_2}^{'}}.
\end{equation}
This may be expressed as
\begin{equation}
\frac{{T}^*}{T_1} =  \frac{2 \sqrt{\xi^*} }{ (1+ \xi^*) } \sqrt{\theta}.
\end{equation}
iv) Fig. 1 shows the behaviour of subsystem and effective temperatures
as function of efficiency. Particularly, $T$ shows a nonmonotonic trend.
For a given value of $\eta$, $T$ in Fig. 1 corresponds to that engine set-up
which yields the maximum work. From numerical calculations, it is observed that the temperature
has a minimum at an efficiency which is bounded from below by Curzon-Ahlborn value.

\begin{figure}[ht]
\vspace{0.2cm}
\includegraphics[width=7cm]{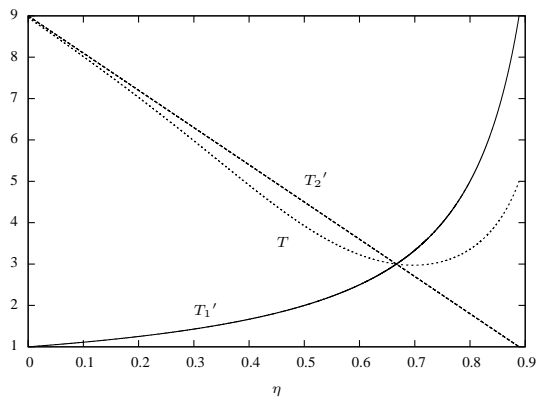}
\caption{ For $T_2 =1$ and $T_1 = 9$, the subsystem temperatures ${{T}_1}^{'}, {{T}_2}^{'}$
and the effective temperature $T$ of the composite system evaluated at maximum work
corresponding to a given efficiency.  
$T$ is given by a weighted average over the subsystem temperatures and so its curve
 lies in between the curves for subsystem temperatures. All
the three temperatures are equal at Curzon-Ahlborn efficiency. }
\hfill
\label{fig1}
\end{figure}
\section{Effective temperatures: a comparison}
The definition of nonequilibrium temperature is not unique for a given situation 
and one can envisage different definitions.
We compare with our defintion two other definitions  of the effective temperature 
from literature, that are relevant to our system. The first candidate is 
the spectral temperature \cite{Mahler2004}. This definition depends only on the
energy probability distribution and the energy spectrum of the system and is
applicable even for nonequilibrium situations. Thus for
a non-degenerate spectrum, the inverse of spectral temperature is defined to be
\begin{equation}
\frac{1}{T_s} = -\left(  1-\frac{P_0 + P_M}{2} \right)^{-1} \sum_{i=1}^{M} \left( \frac{P_i + P_{i-1}}{2}
    \right) \frac{\ln P_i -\ln P_{i-1}  }{E_i-E_{i-1}},
\end{equation}
where Boltzmann's constant has been set to unity. $P_i$ is the probability to
occupy a level with energy $E_i$ and index for the levels ranges from $0$ (ground state) to $M$. 
For our case of two TLS in the final state after work extraction, 
using the values $\{E_i  \}\equiv \{0, a_2, a_1, a_1+a_2 \}$
and $\{ P_i \} \equiv \{r_1s_1, r_2s_1, r_1s_2, r_2 s_2 \}$,
the (inverse) spectral temperature is explicitly given by
\begin{equation}
\frac{1}{T_s} =  \frac{1}{{T}_{1}^{'}} \frac{(\nu-\theta)}{(\nu-\nu^2)}\frac{x}{(1+x)}
 + \frac{1}{{T}_{2}^{'}}    \frac{1}{(1+x)}, 
\end{equation}
where $x = (r_1 + s_1 -2 r_1 s_1)$. The special cases include: Carnot limit,
when $\nu \to \theta$ and so $T_s = (1+x){T}_{2}^{'}$; when $\nu \to 0$,
$T_s \to 0$. Finally, for CA efficiency ($\nu = \sqrt{\theta}$), the spectral temperature is equal to
the subsystems' temperature. In general, the behaviour of $T_s$ as shown in Fig. 2  is quite different
from the proposed definition. 
\begin{figure}[ht]
\vspace{0.2cm}
\includegraphics[width=7cm]{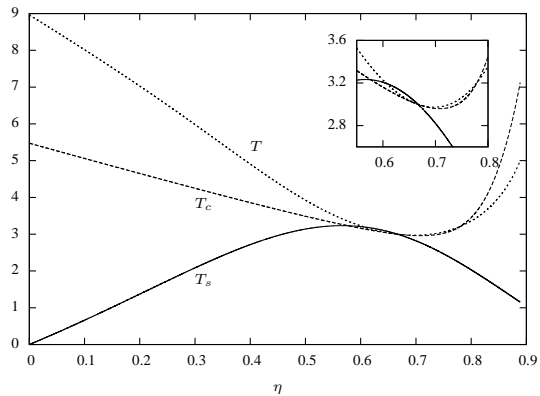}
\caption{ Comparison between other definitions of nonequilibrium temperatures for $T_2 =1$ and $T_1 = 9$. 
$T$ is same as in Fig.1 while $T_s$ denotes the spectral temperature and $T_c$, the contact temperature, as defined
in section IV. Inset shows the enlarged region around the point where all three temperatures 
are equal, which is at $\eta = 1-\sqrt{\theta}= 0.6667$.}
\hfill
\label{fig2}
\end{figure}

The second definition we consider is also called as the contact temperature ($T_c$).
If a general nonequilibrium system whose different parts may be at different
local temperatures, is brought in contact with such a heat bath, that
some parts of the system give heat to the latter and some absorb heat from
 it, so that the net heat transferred between the system and the bath
is zero, then the temperature of that bath defines $T_c$ \cite{Muschik1977}. In other words,
energy conservation holds for the system and different parts of it come
to a common temperature equal to that of the heat bath. Thus for two TLS,
we impose that the total mean energy calculated with canonical distributions
for each TLS, corresponding to a temperature $T_c$,  is equal to the final
mean energy $U^{'} = a_1 s_2 + a_2 r_2$. The temperature obtained numerically
is depicted in Fig. 2. The behaviour of the contact temperature is closer 
to the proposed definition in regions where subsystem temperatures are equal
or nearly to each other. However, towards the extreme values of the engine efficiency,
the two temperatures take on  different values.
\section{Conclusions}
The notion of temperature is well understood in the domain of
equilibrium thermodynamics. However, its extension to nonequilibrium situations
is non-trivial. See for example \cite{Jou2003} for a review of effective temperatures in nonequilibrium
situations. In this paper, we have discussed a quantum heat engine in which two TLS
prepared in different thermal states, undergo a unitary thermally isolated process and
deliver work to an external work source. The final state of the two-TLS system
is passive (i.e. no further work can be extracted from it) but a nonequilibrium state 
where each subsystem may have a different local temperature.
We have proposed a thermodynamic definition to calculate effective temperature of the 
composite system in its final state. The obtained formula is very similar
to the one expected on thermodynamic grounds. The proposed definition
is compared with the spectral temperature, which seems to have a widely
different behaviour. The other definition called contact temperature appears to
have some semblence to our definition. All the three definitions converge
for mutual equilibrium, but at Carnot limit or the vanishing efficiency they differ from each other
significantly. Future experiments on measurement of temperatures in such systems may decide
between the different definitions. Finally, it will be interesting to extend
these ideas to more elaborate models such as involving entanglement between
the TLS \cite{Zhang2008}. It is hoped that the present analysis
will help to understand thermodynamic behaviour revealed by
quantum heat engines. 
\section{ACKNOWLEDGEMENTS}
The author is thankful to Armen Allahverdyan for useful comments on an initial
draft of this paper. Financial support from University Grants Commission, India 
(Grant No. F.6-1(28)/2007(MRP/Sc/NRCB)) is gratefully acknowledged.

\end{document}